\documentstyle[psfig]{aipproc}

\def\la{\mathrel{\hbox{\rlap{\hbox{\lower4pt\hbox{$\sim$}}}\hbox{$<$}}}}
\def\ga{\mathrel{\hbox{\rlap{\hbox{\lower4pt\hbox{$\sim$}}}\hbox{$>$}}}}

\begin{document}
\title{The GRB/SN Connection:  An Improved Spectral Flux Distribution for the SN-Like Component to the Afterglow of GRB 970228, the Non-Detection of a SN-Like Component to the Afterglow of GRB 990510, and GRBs as Beacons to Locate SNe at Redshifts z $\approx$ 4 -- 5}

\author{Daniel E. Reichart$^*$, Donald Q. Lamb$^*$, and Francisco J. Castander$^{\dagger}$} 
\address{$^*$Department of Astronomy \&
Astrophysics, University of Chicago,\\ 5640 South Ellis Avenue, Chicago
IL, 60637\\ $^{\dagger}$Observatoire Midi-Pyr\'en\'ees, 14 Av. Edouard
Belin, 31400 Tolouse, France}

\maketitle

\begin{abstract}    
We better determine the spectral flux distribution of the supernova
candidate associated with GRB 970228 by modeling the spectral flux
distribution of the host galaxy of this burst, fitting this model to
measurements of the host galaxy, and using the fitted model to better
subtract out the contribution of the host galaxy to measurements of the
afterglow of this burst.  Furthermore, we discuss why the non-detection of a SN1998bw-like component to the afterglow of GRB 990510 does not necessarily imply that a SN is not associated with this burst.  Finally, we discuss how bursts can be used as beacons to locate SNe out to redshifts of $z \approx 4 - 5$.
\end{abstract}

\section*{An Improved Spectral Flux Distribution for the SN-like Component to the Afterglow of GRB 970228} 

The discovery of what appear to be SNe dominating the light curves and
spectral flux distributions (SFDs) of the afterglows of GRB 980326
(Bloom et al. 1999) and GRB 970228 (Reichart 1999; Galama et al. 1999)
at late times after these bursts strongly suggests that at least some,
and perhaps all, of the long bursts are related to the deaths of
massive stars.  Here, we build upon the results of Reichart (1999) by
modeling the SFD of the host galaxy of GRB 970228, fitting this
model to measurements of the host galaxy, and using the fitted model
to better subtract out the contribution of the host galaxy to
measurements of the afterglow of this burst.

In Figure 1, we plot the observed SFD of the host galaxy of GRB 970228,
as measured with {\it HST}/WFPC2, {\it HST}/NICMOS2, and Keck I
(Castander \& Lamb 1999a; Fruchter et al. 1999).  To these measurements
and a broadband measurement made with {\it HST}/STIS (Castander \&
Lamb 1999a; Fruchter et al. 1999), which is not plotted, we fit a
two-parameter, spectral synthesis model (see Castander \& Lamb 1999a for
details).  The two parameters are the normalization of the SFD, and the
age of the galaxy, defined to be the length of time that star formation
has been occurring at a constant rate.  Taking
$A_V = 1.09$ mag for the Galactic extinction along the line of sight
(Castander \& Lamb 1999b), we find a fitted age of $270^{+460}_{-180}$
Myr; different values of $A_V$ affect primarily the fitted age, and not
the fitted SFD.  Furthermore, models in which star formation slows
considerably, or ceases, are generally too red to account
for the measurements.  Finally, we note that the fitted J- and R-band
spectral fluxes are perfectly consistent with what one finds simply
from linear interpolation between adjacent photometric bands. 

\begin{figure}
\begin{minipage}[t]{2.75truein}
\mbox{}\\
\psfig{file=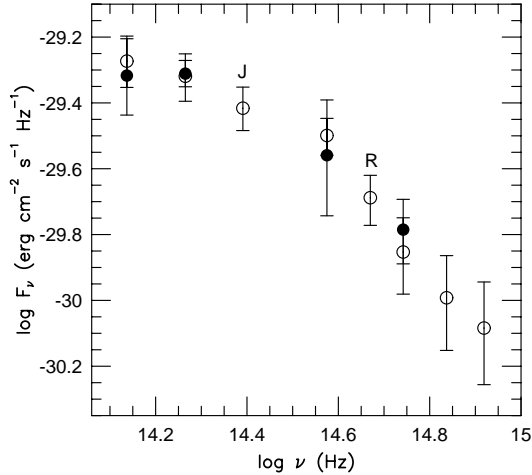,width=2.75truein,clip=}
\end{minipage}
\hfill
\begin{minipage}[t]{2.75truein}
\mbox{}\\
\caption{The observed (filled circles) and modeled (unfilled circles) K- through U-band SFDs of the host galaxy of GRB 970228.}
\end{minipage}
\end{figure}


In Figure 2 (left panel), we plot the SFD of the afterglow minus the
{\it observed} SFD of the host galaxy from Figure 1.  For the SFD of
the afterglow, we use the revised K-, J-, and R-band measurements of
Galama et al. (1999) and the I- and V-band measurements of Castander \&
Lamb (1999a; see also Fruchter et al. 1999); all of these measurements
were taken between 30 and 38 days after the burst.  We have scaled
these measurements to a common time of 35 days after the burst, and
have corrected these measurements for Galactic extinction along the
line of sight (see Reichart 1999 for details).  The K-band measurement
of the afterglow is consistent with that of the host galaxy (Galama et
al. 1999), resulting in an upper limit in Figure 2; J- and R-band
measurements of the host galaxy are not available, again resulting in
upper limits in Figure 2.  As originally concluded by Reichart (1999),
this SFD is consistent with that of SN 1998bw, after transforming it to
the redshift of the burst, $z = 0.695$ (Djorgovski et al. 1999), and
correcting it for Galactic extinction along its line of sight (see
Reichart 1999 for details).

\begin{figure}[t]
\psfig{file=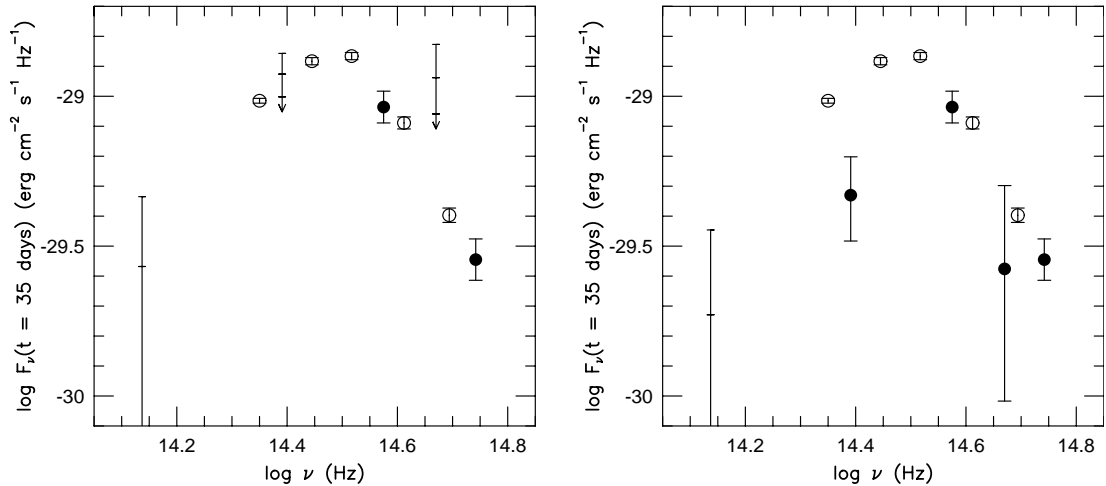,width=5.75truein,clip=} 
\caption{The K- through V-band SFDs of the late afterglow of GRB 970228 after subtracting out the observed (left panel) and modeled (right panel) SFDs of the host galaxy from Figure 1 and correcting for Galactic extinction (filled circles and upper limits), and the I- through U-band SFD of SN 1998bw after transforming to the redshift of GRB 970228, $z = 0.695$, and correcting for Galactic extinction (unfilled circles).  The K-, J-, and R-band upper limits are 1, 2, and 3 $\sigma$.}
\end{figure}

In Figure 2 (right panel), we plot the same distribution, but minus the
{\it modeled} SFD of the host galaxy from Figure 1.  The SN-like
component to the afterglow is detected in the J band, and possibly in
the R band.  The J-band measurement suggests that the SN-like component
is $\approx 1/2$ mag fainter, and $\approx 1/2$ of a photometric band
bluer, than SN 1998bw; however, this difference in J-band spectral
fluxes is significant only at the $\approx 2.5$ $\sigma$ level.  When
possible photometric zero point errors and uncertainties in our
spectral synthesis model of the SFD of the host galaxy are included,
this difference is significant only at the $\approx 2$ $\sigma$ level. 
However, it is suggestive of what is generally expected:  the Type Ic
SNe that are theorized to be associated with bursts (e.g., Woosley
1993) are not expected to be standard candles.

\section*{The Non-Detection of a SN-like Component to the Afterglow of GRB 990510}

Given the rapid rate at which the afterglow of GRB 990510 faded,
$t^{-2.4}$ (Stanek et al. 1999) or $t^{-2.2}$ (Harrison et al. 1999),
at late times after the burst, a SN1998bw-like component to the
afterglow, if present, could have dominated the light curve after about
a month at red and NIR wavelengths (Figure 3; Lamb \& Reichart 1999). 
However, Fruchter et al. (1999b) find no evidence of such a component,
to within a factor of 3 -- 7 in brightness, from two broadband {\it
HST}/STIS observations.  

Caution is in order before one concludes that a SN was not associated
with GRB 990510, or one uses the non-detection of SN1998bw-like
components to afterglows to place lower limits on the redshifts of
bursts.  In the case of GRB 990510, the above calculation requires that
assumptions be made about (1) the form of the SFD of the afterglow at
the times of the observations, since the observations spanned a
wavelength range of 300 -- 900 nm; (2) how to extrapolate the
non-power-law light curve of the early afterglow to the times of the
observations; e.g., Stanek et al. (1999) and Harrison et al. (1999) do
this differently; (3) the range of the luminosity distribution of SNe
associated with bursts, relative to the luminosity of SN 1998bw, since
these SNe are not expected to be standard candles; (4) the SFD of SN
1998bw at ultraviolet wavelengths, since the redshift of this burst is
$z = 1.619$ (Vreeswijk et al. 1999); (5) whether differences in host
galaxy extinction along the SN 1998bw and GRB 990510 lines of sight can
be ignored; and (6) the underlying cosmological model.

\begin{figure}
\begin{minipage}[t]{2.75truein}
\mbox{}\\
\psfig{file=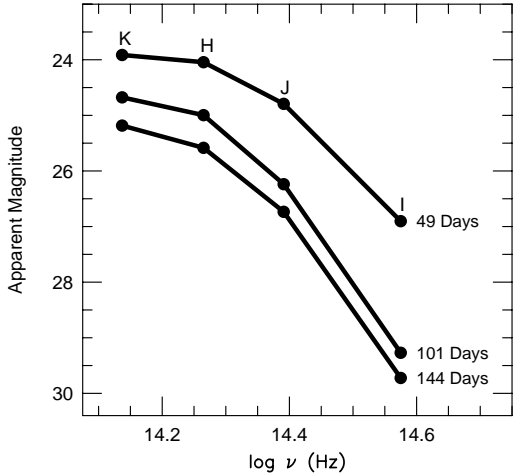,width=2.75truein,clip=}
\end{minipage}
\hfill
\begin{minipage}[t]{2.75truein}
\mbox{}\\
\caption{The SFD of SN 1998bw, transformed to the redshift of GRB 990510, $z = 1.619$, and corrected for Galactic extinction along the line of site (see Lamb \& Reichart 1999 for details).  Had NICMOS not run out of cryogen half a year earlier, the presence or absence of a SN component to the afterglow of this burst could have been firmly established.}
\end{minipage}
\end{figure}

\section*{GRBs as Beacons to Locate Supernovae at Very High Redshifts}

If bursts are indeed associated with SNe, then the first bursts should
have occurred shortly after the first stars formed, at redshifts of $z
\approx 15 - 20$ (Ostriker \& Gnedin 1996; Gnedin \& Ostriker 1997;
Valageas \& Silk 1999).  Lamb \& Reichart (1999) show that bursts and
their afterglows should be detectable out to these very high
redshifts.  One way, of many (see Lamb \& Reichart 1999 for details),
in which bursts can be used to probe the early universe is to use them
as beacons to locate SNe at very high redshifts.  

In Figure 4, we plot the V-band light curve of SN 1998bw added to the
best-fit source-frame V-band light curve of the early afterglow of GRB
970228 from Reichart (1999), transformed to various redshifts and
corrected for Galactic extinction (see Lamb \& Reichart 1999 for
details).  We use the V band because SN 1998bw peaked in this band.
Clearly, the peak of the SN component of the light curve can be
detected, at least from space, out to the K band, which corresponds to
a redshift of $z \approx 3$.  The peak cannot be detected at longer
wavelengths since such faint magnitudes cannot be reached in the L and M
bands.  However, from the K band, one should be able to detect SNe like SN 1998bw blueward of the peak out to redshifts of $z \approx 4 - 5$.

\begin{figure}[t]
\psfig{file=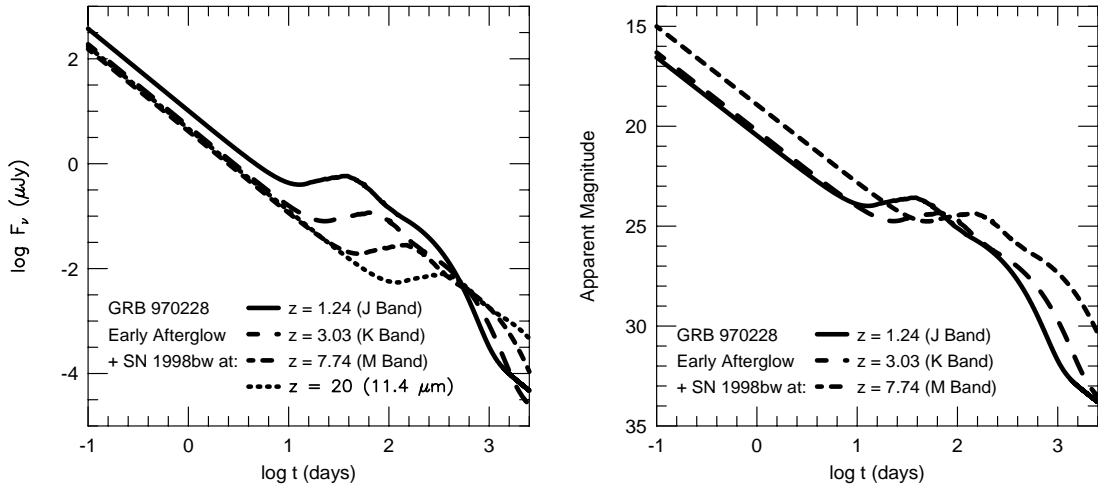,width=5.75truein,clip=}
\caption{V-band light curve of SN 1998bw added to the best-fit source-frame V-band light curve of the early afterglow of GRB 970228 from Reichart (1999), transformed to various redshifts and corrected for Galactic extinction}
\end{figure}

\end{document}